\documentclass{elsarticle}
\usepackage{graphicx}
\makeatletter

\def\ps@pprintTitle{%

   \let\@oddhead\@empty

   \let\@evenhead\@empty

 \let\@oddfoot\@empty

  \let\@evenfoot\@oddfoot
}
\makeatother

\def\ee{\end{eqnarray*}}
\def\be{\begin{eqnarray*}}
\def\bee{\end{eqnarray}}
\def\bbe{\begin{eqnarray}}

\def\n{\mathbf n}
\def\N{\mathbf N}

\def\t{\mathbf t}
\def\p{\partial }
\def\q{\mathbf q}

\def\pp{\mathbf p}

\begin{document}
\title{Ill-posedness of  waterline integral of time domain free surface Green function  for   surface piercing body advancing at dynamic speed}

\author{Zhi-Min Chen}
\address{
School of Mathematics and Statistics, Shenzhen University, Shenzhen, 518052,
China}

\begin{abstract}

In the linear time domain computation of a floating body advancing at a dynamic speed, the source formulation   for the velocity potential of the hydrodynamic  problem is commonly used so that the velocity potential is expressed as  the integral of time domain free surface sources distributed on the two-dimensional  wetted body surface and the one-dimensional waterline, which is the intersection of the wetted body surface and the mean free water surface. A time domain free surface source is corresponding to the time domain free surface Green function associated with a suitable source strength, which is to be solved from  body boundary condition and normal velocity boundary integral equation  of the source formulation.

The normal velocity boundary integral equation contains an integral of the normal derivative of the time domain free surface Green function on the waterline. It is shown that the waterline integral is ill-posed. Thus the source strength of velocity potential is not obtainable.

\


\end{abstract}
\maketitle

\section{Introduction} With the rapid development of computing capacity, the dynamic behaviour of a ship advancing in waves has been increasingly predicted  in satisfactory manners. Nonlinear time domain methods are more and more displacing linear time domain methods in numerical simulations.  However, if we consider fluid motion problem  from the viewpoint of rigorous analysis, the situation is quite different.
It is the purpose of the present paper to show a troublesome in a linear time domain method with respect to  the integral on  the  waterline contour, the intersection between the ship surface and the mean water surface.

 As a fundamental formulation in hydrodynamics, the  velocity potential of  the wave-body motion problem is expressed as an  integral  of the time domain free surface sources distributed on  the mean wetted body surface and  the waterline  contour. The time domain  free surface Green function at a source point  represents the potential of a free surface source with the unit strength.

From \cite{Beck,BM,Has,We}, the time domain free surface Green function with respect to  a field point $\q=(x,y,z)$   and a source point $\pp=(\xi,\eta,\zeta)$    is expressed as
\bbe \label{abc}\hat G(\q,\pp,t-\tau)=\frac{\delta(t-\tau )}{|\q-\pp|}-\frac{\delta(t-\tau )}{|\q-\bar \pp |}+  H(t-\tau) G(\q,\pp,t-\tau).
\bee
Here   $\bar \pp =(\xi,\eta,-\zeta)$ is the image of  $\pp$, $H$ is the Heaviside step function,  $\delta$ is the Dirac delta function and  $G$ is  the wave integral
\bbe \label{n1}  G=2 \int^\infty_{0}\sqrt{gk}  \sin \left((t-\tau)\sqrt{kg}\right)  e^{k(z+\zeta)} J_0(k|(x,y)-(\xi,\eta)|)dk,
\bee
where $g$ is the gravitational acceleration   and $J_0$ the zero order Bessel function of the first kind.

For simplicity, we consider a surface piercing body advancing  at a dynamic speed $U=U(t)>0$ in  the positive  $Ox$ direction. This gives rise to the dynamic fluid domain $\mathcal{D}(t)$ and the mean wetted body surface $S_B(t)$ upper bounded by the linear free surface $z=0$.
The governing equation of the fluid motion problem for the velocity perturbation potential $\phi$ is  the Laplace equation
\bbe &&\Delta \phi =0\label{z1}  \ \ \mbox{ in  $\mathcal{D}(t)$}.
\bee
The function $\phi$ can be evaluated uniquely due to  the linear free surface boundary condition
\bbe\label{z2}
&&\frac{\partial^2\phi}{\partial t^2}+g \frac{\partial \phi}{\partial z}=0 \ \ \  \mbox{ on the  mean free surface } z=0,
\bee
the initial value  condition
\bbe\label{z3}
&&\phi|_{t=0}=0,\,\,\, \left.\frac{\partial \phi}{\partial t}\right|_{t=0}=0,
\bee
the boundary condition at the infinity
\bbe
\lim_{\q \in \mathcal{D}(t), \ |\q|\to \infty }\nabla \phi = 0,\label{DDD}
\bee
 and the body  boundary condition
\bbe
\frac{\p \phi}{\p \n} = (U,0,0)\cdot \n \ \ \ \mbox{ on  $S_B(t)$,}\label{body}
\bee
where $\n=(n_1,n_2,n_3)$ denotes the unit normal vector field of $S_B(t)$ pointing into the body.

As in  \cite{Lia,Ma}  in the extension of the Brard \cite{Bra} waterline  integral,  the application of Green's theorem and the conditions (\ref{z1})-(\ref{DDD}) implies that the velocity potential $\phi$ of the fluid motion  satisfies the following boundary integral equation
\bbe \label{ann2}
&&4\pi\phi(\q,t) +  \int^t_0d\tau \int_{S_B(\tau)} \left(\phi(\pp,\tau)\frac{\partial \hat G(\q,\pp,t-\tau)}{\partial \n_\pp}-\hat G(\q,\pp,t-\tau) \frac{\partial \phi(\pp,\tau)}{\partial \n_\pp}\right)dS\nonumber
\\
&&=-\frac1g\int^t_0 d\tau\oint_{\Gamma(\tau)} \left( \phi(\pp,\tau) \frac{\partial G(\q,\pp,t-\tau)}{\partial \tau}-G(\q,\pp,t-\tau)\frac{\partial \phi(\pp,\tau)}{\partial \tau}\right)U_\N (\pp,\tau) dS
\label{line}
\bee
for a field point $\q\in \mathcal{D}(t)$,  the waterline contour $\Gamma(t)=S_B(t) \cap \{ \pp=(\xi,\eta,\zeta); \zeta=0\}$ and  the two-dimensional normal velocity
$U_\N=(U(\tau),0)\cdot (n_1,n_2)$ of $\Gamma(\tau)$.  The contour   integral is in the anticlockwise direction. As shown in (\ref{line}), the integrand of the  waterline integral involves the time derivative $\p_\tau \phi$ which leads to  difficulty in computing $\phi$ from the integral equation (\ref{line}). However, it was suggested by Beck and Magee  \cite{BM} that it is convenient to use the source formulation of (\ref{line}) expressed  in the following form
\bbe \nonumber
&&4\pi\phi(\q,t) +  \int_{S_B(t)} \left( \frac{\sigma(\pp,t)}{|\q-\pp|}-\frac{\sigma(\pp,t)}{|\q-\bar \pp |}\right) dS+\int^t_0d\tau \int_{S_B(\tau)}\sigma(\pp,\tau)  G(\q,\pp,t-\tau) dS
\\&&\hspace{10mm}=\frac1g\int^t_0 d\tau\oint_{\Gamma(\tau)} \sigma(\pp,\tau) G(\q,\pp,t-\tau)U_\n(\pp,\tau) U_\N(\pp,\tau)dS\label{nn333}
\bee
for field point in the fluid domain  $\q\in \mathcal{D}(t)$,  the unknown source strength $\sigma$ and the three-dimensional normal velocity $U_\n= (U,0,0)\cdot \n$ of $\Gamma(\tau)$. This  source formulation  (see  \cite{Ma} for details) is derived from (\ref{line}) together with the existence of  the Dirichlet  problem  as in Lamb \cite{Lamb}. Applying the normal derivative to (\ref{nn333}) with $\q$ moving to the body boundary $S_B$, we may obtain the following boundary integral equation
\bbe \nonumber
&&4\pi\frac{\partial \phi(\q,t) }{\partial \n}+ \lim_{\hat \q\in \mathcal{D}, \hat \q \to \q}\int_{S_B(t)}\sigma(\pp,t) \frac{\partial }{\partial \n_\q}\left(\frac1{|\hat\q-\pp|}-\frac1{|\hat\q-\bar \pp |}\right) dS
\\
&&+\int^t_0d\tau \int_{S_B(\tau)}\sigma(\pp,\tau) \frac{\partial  G(\q,\pp,t-\tau)}{\partial \n_\q} dS\nonumber
\\&&\hspace{10mm}=\frac1g\int^t_0 d\tau\oint_{\Gamma(\tau)} \sigma(\pp,\tau)\frac{\partial G(\q,\pp,t-\tau)}{\partial \n_\q}U_\n(\pp,\tau) U_\N(\pp,\tau)dS.\label{nn44}
\bee
 The source strength $\sigma$ is determined by body boundary condition (\ref{body}) and the  boundary integral equation (\ref{nn44}).

When $\q \in \Gamma(t)$, the Rankine Green function part $\frac1{|\q-\pp|}-\frac1{|\q-\bar \pp |}=0$ and so  (\ref{nn44}) reduces to
\bbe \nonumber
&&4\pi\frac{\partial \phi(\q,t) }{\partial \n}+\int^t_0d\tau \int_{S_B(\tau)}\sigma(\pp,\tau) \frac{\partial  G(\q,\pp,t-\tau)}{\partial \n} dS\nonumber
\\&&\hspace{10mm}=\frac1g\int^t_0 d\tau\oint_{\Gamma(\tau)} \sigma(\pp,\tau)\frac{\partial G(\q,\pp,t-\tau)}{\partial \n}U_\n(\pp,\tau) U_\N(\pp,\tau)dS.\label{nn4444}
\bee
Thus the source strength $\sigma$ on the waterline $\Gamma$ is determined by (\ref{nn4444}) associated with the body boundary condition (\ref{body}).

Therefore to evaluate the velocity potential $\phi$ of  the linear hydrodynamics problem (\ref{z1})-(\ref{body}) (see, for example, \cite{Beck,BM,Lia,Ma}), one has   to provide  Green function  approximation \cite{CQP,Cl,DD,Liu1,Ma,N} of $G$ and then derive the source strength $\sigma$ from the combination of the body boundary condition (\ref{body}) and the body boundary integral equations  (\ref{nn44}) and (\ref{nn4444}) with respect  the field point $\q\in S_B(t) \cup \Gamma(t)$.  To do so, it is necessary to evaluate the waterline  integral involved in the boundary integral equation (\ref{nn4444}).
The difficulty on the  numeral evaluation of the waterline integral  is well known  but the   rigorously   analytic discussion on the integral is missing.  In contrast to  earlier investigations  based on well-posedness assumption  of the boundary integral equation,
we  show the waterline integral on the right-hand side of (\ref{nn4444}) is unbounded around a field point $\q \in \Gamma(t)$ whenever the body surface  $S_B(t)$ at $\q$ is not perpendicular to  the mean free surface $z=0$.

The wave-body motion problem can also be modelled mathematically by using frequency domain free surface Green functions if the body  is advancing at a uniform speed. Therefore  the velocity potential of the fluid motion problem can be derived numerically  based on the numerical evaluation of the frequency domain free surface Green function  (see \cite{Chen2012,Chen2014b}).

Now we begin with the understanding of the Green  function on the free surface.

\section{The Green function on the waterline}
The wave integral $G$ of the Green function is harmonic  and hence is smooth in the fluid domain $z<0$. However this smoothness property is not extendible to the free surface $z=0$. The analytical behaviour of $G$ in $z<0$ is quite different to that on $z=0$.
 For the limit situation when both field and source points are in the mean  free surface or $z=\zeta=0$, the wave integral reduces formally to
\bbe \label{nn1}  G&=& 2\int^\infty_{0}\sqrt{gk}  \sin \left((t-\tau )\sqrt{kg}\right)   J_0(kR)dk
\\
&=&  2\sqrt{\frac g{R^3}} \int^\infty_{0}\sqrt{\lambda}  \sin \left(s\sqrt{\lambda}\right)   J_0(\lambda)d\lambda
\bee
for  $R=|(x,y)-(\xi,\eta)|$, $\lambda = kR$ and $s= (t-\tau )\sqrt{\frac gR}$.

It has been  displayed in \cite[Equations. (22.18), (22.19) and (22.21)]{We} that the dimensionless wave integral  can be expressed by using Bessel functions of the first kind:
\bbe\label{ab3}
\int^\infty_{0}\sqrt{\lambda}  \sin \left(s\sqrt{\lambda}\right)   J_0(\lambda)d\lambda
= \frac{\pi s^3}{16\sqrt{2}}\left(J_{\frac14}\left(\frac{s^2}8\right)J_{-\frac14}\left(\frac{s^2}8\right)+J_{\frac34}\left(\frac{s^2}8\right)J_{-\frac34}\left(\frac{s^2}8\right)\right).
\bee
This, together with the asymptotic expression \cite{A}
\bbe J_\alpha(\lambda )\sim  \sqrt{\frac 2{\pi \lambda}}\cos \left(\lambda-\frac{\alpha \pi}2-\frac\pi 4\right) \ \mbox{ for large } \lambda >0,\label{aab3}
\bee
implies the asymptotic expression
\bbe\label{aaax}
\int^\infty_{0}\sqrt{\lambda}  \sin \left(s\sqrt{\lambda}\right)   J_0(\lambda)d\lambda \sim
\frac{s}{\sqrt{2}}\sin\left(\frac{s^2}4\right)\ \mbox{ for large } s >0 
\bee
or
\bbe\label{G}
G\sim \sqrt{2} \frac{g(t-\tau )}{R^2}\sin \left(\frac{(t-\tau )^2}{4}\frac gR\right) \ \mbox{ for large } t-\tau >0.
\bee
The comparison between the left-hand and right-hand sides  of (\ref{aaax}) is presented in Figure 1, which shows a good agreement except a small deviation around $s=3$.
\begin{figure}\label{Fig11}

 \hspace{0mm}\includegraphics[width = 440pt, height = 120pt]{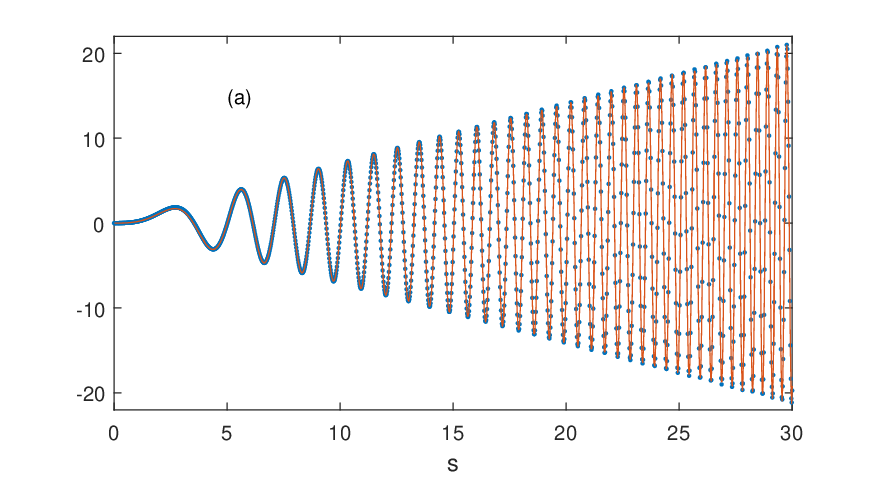}\vspace{-3mm}

  \hspace{-0mm}\includegraphics[width = 440pt, height = 120pt]{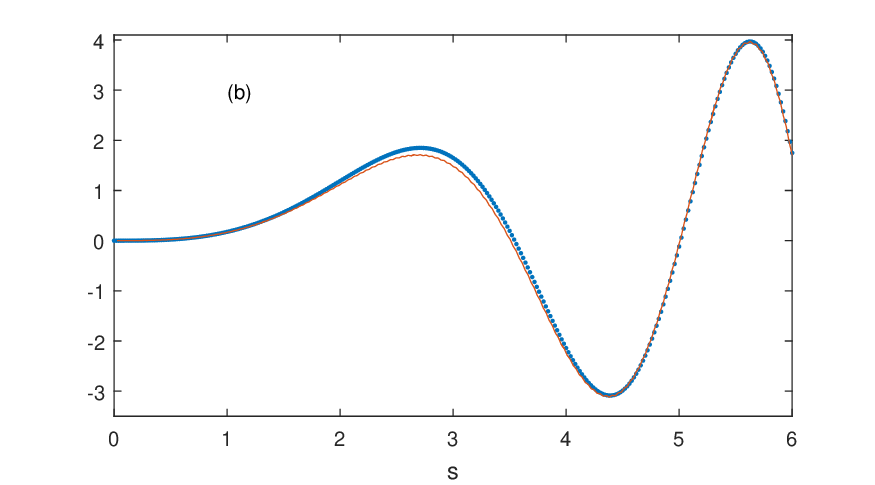}\vspace{-3mm}

  \caption{Comparison  between the numerical  dimensionless wave integral $ \int^\infty_{0}\sqrt{\lambda}  \sin \left(s\sqrt{\lambda}\right)   J_0(\lambda)d\lambda$ (solid lines) and its analytic asymptotic expression  $\frac s{\sqrt{2}}\sin(\frac{s^2}4)$ (points lines) when for $0<s<30$ (a) and $0<s<6$ (b).}
\end{figure}

In this paper, we need the wave integral for small $s>0$. Although Figure 1 shows  that the wave integral on the right-hand side of (\ref{aaax}) agrees well with the analytic function on the right-hand side of (\ref{aaax}) for small $s$, we would like to use an alternative expression of the wave integral by using
the following Taylor expansion \cite[Equation 9.1.14]{A} of the the Bessel functions multiplication
\bbe J_\alpha(\lambda)J_{-\alpha}(\lambda)= \sum_{k=0}^\infty \frac{(-1)^k (2k)!(\frac\lambda2)^{2k}}{\Gamma(k+\alpha+1)\Gamma(k-\alpha+1) (k!)^2}.
\bee
The combination of the previous  formula and the wave integral (\ref{ab3}) yields the Taylor expansion
\be\label{ab33}
 \frac{G}{2\sqrt{\frac{g}{R^3}}}&=&\frac{\pi s^3}{16\sqrt{2}}\left( J_{\frac14}\left(\frac{s^2}8\right)J_{-\frac14}\left(\frac{s^2}8\right)+J_{\frac34}\left(\frac{s^2}8\right)
J_{-\frac34}\left(\frac{s^2}8\right)\right)
\\
&=& \frac{\pi s^3}{16\sqrt{2}}\sum_{k=0}^\infty \frac{1}{ (k!)^2}\left(\frac{(-1)^k (2k)!(\frac{s^2}{16})^{2k}}{\Gamma(k+\frac14+1)\Gamma(k-\frac14+1)}+\frac{(-1)^k (2k)!(\frac{s^2}{16})^{2k}}{\Gamma(k+\frac34+1)\Gamma(k-\frac34+1) }\right)
\ee
and hence
\bbe \label{nn1x}G=  \frac{\pi g^2}{8\sqrt{2}}\sum_{k=0}^\infty \frac{(t-\tau)^{4k+3}}{R^{2k+3} (k!)^2}\left(\frac{(-1)^k (2k)!(\frac{g}{16})^{2k}}{\Gamma(k+\frac14+1)\Gamma(k-\frac14+1)}+\frac{(-1)^k (2k)!(\frac{g}{16})^{2k}}{\Gamma(k+\frac34+1)\Gamma(k-\frac34+1) }\right).
\bee
This gives rise to  the following partial derivatives
\bbe\label{Gtt}
\frac{\partial^2 G}{\partial t^2}
 = \frac{\pi g^2}{8\sqrt{2}}\sum_{k=0}^\infty \frac{(4k+3)(4k+2)(t-\tau)^{4k+1}}{ R^{2k+3} (k!)^2}\left(\frac{(-1)^k (2k)!(\frac{g}{16})^{2k}}{\Gamma(k+\frac54)\Gamma(k+\frac34)}+\frac{(-1)^k (2k)!(\frac{g}{16})^{2k}}{\Gamma(k+\frac74)\Gamma(k+\frac14) }\right)
\bee
and
\bbe \label{GR}\frac{\p G}{\p R}=-\frac{\pi g^2}{8\sqrt{2}}\sum_{k=0}^\infty \frac{(2k+3)(t-\tau)^{4k+3}}{ R^{2k+4}(k!)^2}\left(\frac{(-1)^k (2k)!(\frac{g}{16})^{2k}}{\Gamma(k+\frac54)\Gamma(k+\frac34)}+\frac{(-1)^k (2k)!(\frac{g}{16})^{2k}}{\Gamma(k+\frac74)\Gamma(k+\frac14) }\right)
\bee
to be used in the analysis on  the waterline integral.

\section{Ill-posedness}
For the surface piercing body advancing at a continuous dynamic speed $U=U(t)>0$  in the positive $Ox$ direction,  the waterline  at $\tau$ is expressed  as
\bbe \label{f00}\Gamma(\tau) = \Gamma(0) + (\int^\tau_0 U(s)ds,0,0).
\bee
For a field point $\q \in \Gamma(t)$ initially from $\q_0\in \Gamma(0)$, the wave integral $G(\q,\pp,t,\tau)$
is continuous with respect to $(\pp,\tau)$ except at the  singular point  $(\pp,\tau)=(\q,t)$. For  the unboundedness of  the waterline integral on a small panel $[\pp',\pp'']\times [t-\epsilon,t]$ covering the singular point $(\q,t)$, the dynamic behaviour of the panel elements are specified as
\be
 &&\q=\q_0+(\int^t_0U(s)ds,0,0)\in \Gamma(t),
\\
&&\pp{'}=\q_0+ \t \epsilon +(\int^\tau_0U(s)ds,0,0) \in \Gamma(\tau),
\\
&&\pp{''}=\q_0-\t \epsilon+(\int^\tau_0U(s)ds,0,0)\in \Gamma(\tau),
\\
&&\pp=\q_0+\t l+(\int^\tau_0U(s)ds,0,0)\in \Gamma(\tau),
\\
&&\t=(t_1,t_2,0) - \mbox{ the unit tangential vector of $\Gamma(0)$ at $\q_0$}
\ee
for $t>\tau>t-\epsilon >0$ and $-\epsilon <l<\epsilon$.
Assume the tangential component   $t_2\neq 0$ and  $\Gamma(0)$  being  smooth at $\q_0$  so that the small panel of $\Gamma(0)$ can be approximated by a small segment.
The  waterline  integral on the right hand side of (\ref{nn44}) restricted the the single panel part is
\bbe\nonumber
&&\frac1g\int^t_{t-\epsilon}\int^{\pp' }_{\pp''} \sigma(\pp,\tau)\frac{\partial G(\q,\pp,t-\tau)}{\partial \n_\q}U_\n(\pp,\tau)U_\N(\pp,\tau)dS d\tau
\\
&&=\frac1g\int^t_{t-\epsilon}\int^{\epsilon }_{-\epsilon} \sigma(\pp,\tau)\frac{\partial G(\q,\pp,t-\tau)}{\partial \n_\q}U_\n(\pp,\tau)U_\N(\pp,\tau)dl d\tau\nonumber
\\&&=\frac{\sigma(\q,t)(n_1U)^2}g\int^t_{t-\epsilon}\int^{\epsilon }_{-\epsilon} \frac{\partial G(\q,\pp,t-\tau)}{\partial \n_\q}dl d\tau,
\bee
after the  use of the smallness assumption of $\epsilon>0$ and the continuity of $\sigma$ and $U$.
It suffices to show the unboundedness of  the panel integral of
$\frac{\partial G}{\partial \n}:$
\bbe\nonumber
&&\int^t_{t-\epsilon} \int^{\epsilon }_{-\epsilon} \frac{\partial G(\q,\pp,t-\tau)}{\partial \n_\q}dl d\tau
\\ \nonumber
&&=\int^t_{t-\epsilon} \int^{\epsilon }_{-\epsilon}\left( (n_1,n_2,0)\cdot \frac{\q-\pp}{R}\frac{\partial G(\q,\pp,t-\tau)}{\partial R}+ n_3\frac{\partial G(\q,\pp,t-\tau)}{\partial z}\right) dl d\tau
\\
&&=\int^t_{t-\epsilon} \int^{\epsilon }_{-\epsilon}(n_1,n_2,0)\cdot \frac{\q-\pp}{R}\frac{\partial G(\q,\pp,t-\tau)}{\partial R}dld\tau-\frac{n_3}g\int^t_{t-\epsilon} \int^{\epsilon }_{-\epsilon}\frac{\partial ^2 G(\q,\pp,t-\tau)}{\partial t^2} dld\tau\hspace{10mm}\label{f1}
\bee
due to  the free surface boundary condition (\ref{z2}) of the Green function.

Upon the observation \bbe \frac1R&=&\frac1{|\q-\pp|}
=\frac1{\sqrt{(t_2l)^2+\left(-t_1l +\int^t_\tau U(s)ds \right)^2}},
\label{e0}
\bee
we find that
\be \frac{1}{R}&
\leq& \frac{2}{\int^t_\tau U(s)ds} \ \ \ \mbox{ when } \ \ |t_1l|<\frac12\int^t_\tau U(s)ds
\ee
and
\be \frac{1}{R}&
\leq&\frac{2|t_1|}{|t_2|\int^t_\tau U(s)ds} \ \ \ \mbox{ when } \ \ |t_1l|\ge\frac12\int^t_\tau U(s)ds
\ee
lead to
\bbe \label{e10} \frac{1}{R}&
\leq&2\left(1+\frac{|t_1|}{|t_2|} \right)\frac{1}{\int^t_\tau U(s)ds}
\leq \frac2{U_-}\left(1+\frac{|t_1|}{|t_2|} \right)\frac{1}{t-\tau}
\bee
for $U_-= \min_{t-\epsilon \leq s\leq t}U(s)$.
This implies,  for small $t-\tau>0$,
\bbe \label{e1} \frac{(t-\tau)^2}{R}&
\leq&\frac2{U_-}\left(1+\frac{|t_1|}{|t_2|} \right)(t-\tau)= O(t-\tau).
\bee
Therefore, the application of the estimate  (\ref{e1}), the Gamma function identity \cite[Equation 6.1.31]{A}  and the smallness of the quantity $t-\tau$  into  the partial derivatives (\ref{Gtt}) and (\ref{GR}) yields
\bbe
\frac{\partial^2 G}{\partial t^2}\nonumber
&=&  \frac{3\pi g^2(t-\tau)}{4\sqrt{2}R^3} \left(\frac{ 1}{\Gamma(\frac14+1)\Gamma(1-\frac14)}+\frac{1}{\Gamma(\frac34+1)\Gamma(1-\frac34) }+O((t-\tau)^2)\right)
\\
&=&   \frac{3\pi g^2(t-\tau)}{4\sqrt{2}R^3}\left(\frac{ \sin \frac\pi4}{\frac\pi4}+\frac{\sin\frac{3\pi}4}{\frac{3\pi}4}+O((t-\tau)^2)\right)\nonumber
\\
&=&  \frac {2g^2(t-\tau)}{R^3} +O(1)\label{Gttt}
\bee
and
\bbe \frac{\p G}{\p R}\nonumber
&=&   -\frac{3\pi g^2(t-\tau)^3}{8\sqrt{2}R^4}  \left(\frac{1}{\Gamma(\frac14+1)\Gamma(-\frac14+1)}+\frac{1}{\Gamma(\frac34+1)\Gamma(-\frac34+1) }+O((t-\tau)^2)\right)
\\ \nonumber
&=&   -\frac{3\pi g^2(t-\tau)^3}{8\sqrt{2}R^4}  \left(\frac{ \sin \frac\pi4}{\frac\pi4}+\frac{\sin\frac{3\pi}4}{\frac{3\pi}4}+O((t-\tau)^2)\right)
\\
&=&-\frac{g^2(t-\tau)^3}{R^4} +O(t-\tau).\label{GRR}
\bee

 Let us begin with the estimate of the panel integral of the  partial derivative $\frac{\p^2 G}{\p t^2}$.
Note that
\be dR\leq | dR| =\frac{|(\q-\pp)\cdot(-\t)dl|}{R} \leq dl.\ee
With the use of  (\ref{Gttt}), the half panel integral is calculated as
\be\nonumber
\int^t_{t-\epsilon}\int^{\epsilon}_{0}\frac{\partial ^2 G(\q,\pp,t-\tau)}{\partial t^2} dld\tau
  \nonumber
&&=\int^{t}_{t-\epsilon} d\tau\int^{\epsilon}_{0}2\frac{g^2}{R^3} (t-\tau)dl+O(\epsilon^2)
\\&&\geq \int^{t}_{t-\epsilon} d\tau\int^{\epsilon}_{0}2\frac{g^2}{R^3} (t-\tau)dR+O(\epsilon^2)
\\
&&=\int^{t}_{t-\epsilon}\left[- \frac{g^2(t-\tau)}{|\t \epsilon+\int^t_\tau U(s)ds|^2} + \frac{g^2(t-\tau)}{|\int^t_\tau U(s)ds|^2} \right]d\tau+O(\epsilon^2)
\\
&&\ge\int^{t}_{t-\epsilon}\left[- \frac{g^2(t-\tau)}{|t_2 \epsilon|^2} + \frac{g^2(t-\tau)}{|\int^t_\tau U(s)ds|^2} \right]d\tau+O(\epsilon^2)
\\
&&\ge\int^{t}_{t-\epsilon}\left[- \frac{g^2(t-\tau)}{|t_2 \epsilon|^2} + \frac{g^2}{U_+^2(t-\tau)} \right]d\tau+O(\epsilon^2)
\\
&&=-\frac{g^2}{2|t_2|^2} - \left[\frac{g^2\ln(t-\tau)}{U_+^2} \right]^{\tau=t-0}_{\tau=t-\epsilon}+O(\epsilon^2)
\ee
for $U_+=\max_{t-\epsilon\leq s\leq t} U(s)$.

Similarly, we have the result for another half panel integral
\bbe \int^t_{t-\epsilon}\int_{-\epsilon}^{0}\frac{\partial ^2 G(\q,\pp,t-\tau)}{\partial t^2} dld\tau
  \nonumber
&&=\int^{t}_{t-\epsilon} d\tau\int_{-\epsilon}^{0}2\frac{g^2}{R^3} (t-\tau)dl+O(\epsilon^2)
\\
&&\ge-\frac{g^2}{2|t_2|^2} - \left[\frac{g^2\ln(t-\tau)}{U_+^2} \right]^{\tau=t-0}_{\tau=t-\epsilon}+O(\epsilon^2).
\bee
Hence,  we have the whole panel integral result for the second time derivative of $G$
\bbe\label{f3}
&&\int^t_{t-\epsilon}\int^{\epsilon}_{-\epsilon}\frac{\partial ^2 G(\q,\pp,t-\tau)}{\partial t^2} dld\tau
\ge-\frac{g^2}{|t_2|^2} - \left[\frac{2g^2\ln(t-\tau)}{U_+^2} \right]^{\tau=t-0}_{\tau=t-\epsilon}+O(\epsilon^2).
\bee

On the other hand, observing that
\be \frac{1}{R}&
\leq& \frac1{\sqrt{(t_2l)^2+\frac14(\int^t_\tau U(s) ds)^2}}\leq \frac{1}{\sqrt{|t_2l|\int^t_\tau U(s)ds}} \ \ \ \mbox{ when } \ \ |t_1l|<\frac12\int^t_\tau U(s)ds,
\ee
and
\be \frac{1}{R}&
\leq&\frac1{|t_2l|}\leq \frac{\sqrt{2|t_1|}}{\sqrt{|t_2l|\int^t_\tau U(s)ds}} \ \ \ \mbox{ when } \ \ |t_1l|\ge\frac12\int^t_\tau U(s)ds,
\ee
we have
\bbe \label{e11} \frac{1}{R}&
\leq&\frac{1+\sqrt{2|t_1|}}{\sqrt{|t_2|U_-}} \frac{1}{\sqrt{|l|(t-\tau)}}.
\bee
and hence, by (\ref{e10}),
\bbe \label{e111} \frac{(t-\tau)^3}{R^4}&
\leq&\left(\frac2{U_-}\left(1+\frac{|t_1|}{|t_2|}\right)\right)^3\frac{1+\sqrt{2|t_1|}}{\sqrt{|t_2|U_-}} \frac{1}{\sqrt{|l|(t-\tau)}}.
\bee

Thus by (\ref{GRR}) and (\ref{e111}), the panel integral involving the horizontal derivative $\frac{\p G}{\p R}$
is calculated as
\bbe\nonumber
\lefteqn{\left|\int^t_{t-\epsilon}\int^{\epsilon}_{-\epsilon}(n_1,n_2,0)\cdot \frac{\q-\pp}{R}\frac{\partial G(\q,\pp,t-\tau)}{\partial R} dld\tau\right|}\\
 \nonumber
 &&\leq \int^t_{t-\epsilon}\int^{\epsilon}_{-\epsilon}\left|\frac{\partial G(\q,\pp,t-\tau)}{\partial R} \right| dld\tau
 \\
&&=\int^{t}_{t-\epsilon} d\tau\int^{\epsilon}_{-\epsilon}\left(\frac{g^2(t-\tau)^3}{R^4} +O(t-\tau)\right)dl\nonumber
\\
&&\leq g^2\left(\frac2{U_-}\left(1+\frac{|t_1|}{|t_2|}\right)\right)^3\frac{1+\sqrt{2|t_1|}}{\sqrt{|t_2|U_-}}
\int^{t}_{t-\epsilon} \int^{\epsilon}_{-\epsilon}
  \frac{dld\tau}{\sqrt{(t-\tau)|l|}} \nonumber+O(\epsilon^3)
\\
&&=8\epsilon g^2\left(\frac2{U_-}\left(1+\frac{|t_1|}{|t_2|}\right)\right)^3\frac{1+\sqrt{2|t_1|}}{\sqrt{|t_2|U_-}}
+O(\epsilon^3).
\label{f4}
\bee

Therefore  the combination of (\ref{f1}), (\ref{f3}) and (\ref{f4}) gives the estimate
\bbe\nonumber
&&\left|\int^t_{t-\epsilon} \int^{\epsilon }_{-\epsilon} \frac{\partial G(\q,\pp,t-\tau)}{\partial \n}dld\tau\right|
\\&&\ge  \left| \frac{n_3}g\int^t_{t-\epsilon}\int^{\epsilon}_{-\epsilon}\frac{\partial ^2 G(\q,\pp,t-\tau)}{\partial t^2} dld\tau\right|-\left|\int^t_{t-\epsilon}\int^{\epsilon}_{-\epsilon}(n_1,n_2,0)\cdot \frac{\q-\pp}{R}\frac{\partial G(\q,\pp,t-\tau)}{\partial R}dl d\tau\right|\nonumber
\\
&&\geq -\frac{g|n_3|}{|t_2|^2} - \left[\frac{2g|n_3|}{U_+^2}\ln(t-\tau) \right]^{\tau=t-0}_{\tau=t-\epsilon}+O(\epsilon^2)-8\epsilon  g^2\left(\frac2{U_-}\left(1+\frac{|t_1|}{|t_2|}\right)\right)^3\frac{1+\sqrt{2|t_1|}}{\sqrt{|t_2|U_-}}\nonumber
\\ &&=-\frac{g|n_3|}{|t_2|^2} - \left[\frac{2g|n_3|}{U_+^2}\ln(t-\tau) \right]^{\tau=t-0}_{\tau=t-\epsilon}+O(\epsilon)\label{f5}
\bee
for $\epsilon>0$ sufficiently small.
This implies   the unboundedness of the integral (\ref{f5}) since
\be -\ln(t-\tau)|^{\tau=t-0}_{\tau=t-\epsilon}=\lim_{\tau \to t-0} (-\ln(t-\tau)+\ln\epsilon) =+\infty
\ee
and hence shows the ill-posedness of the boundary integral equation (\ref{nn44}).

\section{Conclusion}

The linearised problem of a surface piercing body advancing at a dynamic speed  is known to be modelled by the  source formulation (\ref{nn333}) defined by   time domain free surface sources  integrated  on the wetted body surface and the waterline  contour (see, for example,  \cite{Ma}).
To find the  velocity potential of the hydrodynamics  problem  one has to evaluate   the  strength $\sigma$  of the sources to be obtained  by the boundary integral equation (\ref{nn44}) together with  normal velocity boundary condition (\ref{body}).

The present study shows that boundary integral equation (\ref{nn44}) is ill-posed since the waterline integral of (\ref{nn44}) is infinite  for the field point $\q$ approaching to the waterline   when the   body surface $S_B$ at $\q$ is not perpendicular to the mean water surface.

It is popular to move the field point $\q$ down from the free water surface  to  approximate    the waterline  integral in earlier studies. However, the ill-posedness of the waterline  integral shows the approximation to be   unrealistic.

\

\noindent \textbf{Acknowledgment} \noindent This  research is partially supported by NSF (grant No. 11571240) of China.


\begin{thebibliography}{99}

\bibitem{A} Abramowitz, M.; Stegun, I. A., eds. (1983) Handbook of Mathematical Functions with Formulas, Graphs, and Mathematical Tables. Applied Mathematics Series. 55. New York: United States Department of Commerce, National Bureau of Standards; Dover Publications.

\bibitem{Beck}  {Beck, R. F.} (1994) Time-domain computations for floating bodies, {Appl. Ocean Res.} {16}, 267-282.


\bibitem{BM}  Beck, R. F.;  Magee, A. R. (1990) Time-domain analysis for
predicting ship motions,   Dynamics of Marine Vehicles and Structures in Waves, Proc. IUTAM symp. at Brunel
University, Elsevier Publishers, Amsterdam,   49-65.

\bibitem{Bra} Brard, R. (1948)  Introduction \`a  l'\'etude th\'eorique du tangage en marche. Bull. Assoc. Tech. Mar. A\'ero. 47, 455-471, Discussion, 472-479.


\bibitem{Chen2012}Chen,  Z. M.   (2012)  A vortex based panel method for potential flow simulation around a hydrofoil, {J. Fluids Strut.} {28}, 378-391.



\bibitem{Chen2014b} Chen, Z. M. (2014) Regular wave integral approach to the prediction of
hydrodynamic performance of submerged spheroid, Wave Motion 51, 193-205.

\bibitem{CQP}
Chuang, J. M.; Qiu, W.;  Peng, H. (2007)
On the evaluation of time-domain Green function, Ocean Engineering 34, 962-969.

\bibitem{Cl} Cl\'ement, A. H. (1998)  An ordinary differential equation for the Green function of
time domain free surface hydrodynamics, J. Engng.  Math. 33, 201-217.


\bibitem{DD}
Duan, W. Y.;  Dai, Y. S. (2001)
New derivation of ordinary differential equations for time-domain free-surface Green Functions,
China Ocean Eng., 15, 499–507



\bibitem{Has} Haskind, M. D. (1946) The oscillation of a ship in still water,  Izv. Akad. Nauk
SSSR. Otd. Tekhn. Nauk, 23-34.





\bibitem{Lamb} Lamb, H. (1932) Hydrodynamics. Cambridge: Cambridge University Press.



\bibitem{Lia}Liapis, S. J. (1986) Time-domain analysis of ship motions, PhD Dissertation, Department of Naval Architecture and Marine
Engineering, University of Michigan, Ann Arbor, Michigan.


\bibitem{Liu1}Liu, C. F.; Teng, B.; Gou, Y.; Sun, L. (2011)
A 3D time-domain method for predicting the wave-induced forces and
motions of a floating body, Ocean Engineering 38, 2142-2150






\bibitem{Ma} Magee, A. R. (1991) Large-amplitude ship motions in the time domain, PhD dissertation, Naval Architecture and Marine Engineering , The University of Michigan, Ann Arbor, Michigan.

\bibitem{N0} Newman, J. N. (1987) Evaluation of the wave-resistance
Green function: Part 1 - The double integral. J. Ship Res.
31. 79-90.


\bibitem{N}Newman, J. N. (1992) The approximation of free-surface Green function, F. Ursell retirement meeting in the University of Manchester, 1990, In: Wave Asymptotic, 
 Cambridge University Press, Cambridge, 107-135.






\bibitem{We} Wehausen, J. V.;  Laitone, E. V. (1960) Surface Waves. In: S. Flugge (eds.), Handbuch Der Physik Vol. 9. New York: Springer,  446-778.


\
\end{thebibliography}
\end{document}